\documentclass[11pt]{article} \textwidth=6in \oddsidemargin=0in
\usepackage{amssymb} 
\begin{document}
\def\ov{\over} \def\be{\begin{equation}} \def\cd{\cdots}
\def\ee{\end{equation}} \def\l{\ell} \def\iy{\infty}
\def\D{\Delta} \def\ep{\epsilon} \def\tl{\tilde}
\def\({\left(} \def\){\right)} \def\k{\kappa} \def\d{\delta} 
\def\e{\eta} \def\x{\xi} \def\inv{^{-1}} \def\L{\Lambda}
\def\ph{\varphi} \def\ps{\psi} \def\Z{\mathbb Z}
\def\ld{\ldots} \def\C{\mathcal C} \def\noi{\noindent}
\def\bc{\begin{center}} \def\ec{\end{center}} \def\qed{\hfill$\Box$}
\def\S{\mathcal S} \def\a{\alpha} \def\b{\beta} 
\def\ch{\chi} \def\g{\gamma} \def\r{\rho} \def\G{\Gamma} 
\def\sp{\vspace{1ex}} \def\Re{{\rm Re}\,} \def\Im{{\rm Im}\,} \def\m{\mu}
\def\M{\mathcal M}  \def\nb{\bar{n}} \def\ab{\bar{a}}  
\def\N{\mathbb N} \def\R{\mathbb R} \def\n{\nu}

\hfill  February 16, 2015

\bc{\large\bf Natural Boundary for a Sum Involving Toeplitz Determinants}\ec

\bc{\large\bf Craig A.~Tracy}\\
{\it Department of Mathematics \\
University of California\\
Davis, CA 95616, USA}\ec

\bc{\large \bf Harold Widom}\\
{\it Department of Mathematics\\
University of California\\
Santa Cruz, CA 95064, USA}\ec

\begin{abstract}
In the theory of the two-dimensional Ising model, the {\it diagonal susceptibility} is equal to a sum involving Toeplitz determinants. In terms of a parameter $k$ the diagonal susceptibility is analytic inside the unit circle, and the authors proved the conjecture that this function has the unit circle as a natural boundary. The symbol of the Toepltiz determinants was a $k$-deformation of one with a single singularity on the unit circle. Here we extend the result, first, to deformations of a larger class of symbols with a single singularity on the unit circle, and then to deformations of (almost) general Fisher-Hartwig symbols.
\end{abstract}

\bc{\bf I. Introduction}\ec

In the theory of the two-dimensional Ising model there is a quantity, depending on a parameter $k$, called the {\it magnetic susceptibility}, which is analytic inside the unit circle. It is an infinite sum over $M,\,N\in\Z$ involving correlations between the spins at sites $(0,0)$ and $(M,N)$. It was shown in \cite{wmtb} to be representable as a sum over $n\ge1$ of $n$-dimensional integrals. In \cite{nickel} B. Nickel found a set of singularities of these integrals which became dense on the unit circle as $n\to\iy$. This led to the (as yet unproved) {\it natural boundary conjecture} that the unit circle is a natural boundary for the susceptibility. 

Subsequently \cite{boukraa} a simpler model was introduced, called the {\it diagonal susceptibility}, in which the sum of correlations was taken over the diagonal sites $(N,N)$. These correlations were equal to Toeplitz determinants, and the diagonal susceptibility was expressible in terms  of a sum involving Toeplitz determinants.

The Toeplitz determinant $D_N(\ph)$ is $\det\,(\ph_{i-j})_{1\le i,j\le N}$, where $\ph_j$ is the $j$th Fourier coefficient of the symbol $\ph$ defined on the unit circle. The sum in question is
\[\sum_{N=1}^\iy[D_N(\ph)-\M^2],\]
where
\[\ph(\x)=\sqrt{{1-k/\x\ov 1-k\,\x}},\]
and $\M$, the spontaneous magnetization, is equal to $(1-k^2)^{1/8}$.
This also (as we explain below) is equal to a sum of $n$-dimensional integrals, the sum is analytic for $|k|<1$, and the singularities of these summands also become dense in the unit circle as $n\to\iy$. This led to a natural boundary conjecture for the diagonal susceptibility, which we proved in \cite{tw2}.  

The question arises whether the occurrence of the natural boundary is a statistical mechanics phenomenon and/or a Toeplitz determinant phenomenon. This note shows that at least the latter is true. We consider here the more general class of symbols
\[\ph(\x)=(1-k\,\x)^{\a_+}\,(1-k\,/\x)^{\a_-}\,\ps(\x),\]
where $\ps$ is a nonzero function analytic in a neighborhood of the unit circle with winding number zero and geometric mean one. We assume $\a_\pm\not\in\Z,\ \Re\a_\pm<1$. The parameter $k$ satisfies $|k|<1$. We define
\be \ch(k)=\sum_{N=1}^\iy[D_N(\ph)-E(\ph)],\label{sum}\ee
where
\[E(\ph)=\lim_{N\to\iy}D_N(\ph).\footnote{When $\ps(\x)=1$ it equals $(1-k^2)^{-\a_+\a_-}$. In general it equals this times a function that extends analytically beyond the unit disc.}\]
Each summand in (\ref{sum}) is analytic in the unit disc $|k|<1$, the only singularities on the boundary being at $k=\pm1$, and the series converges uniformly on compact subsets. Therefore $\ch(k)$ is analytic in the unit disc.
\sp

\noi{\bf Theorem}. The unit circle $|k|=1$ is a natural boundary for $\ch(k)$.
\sp

The result in \cite{tw2} was established, and here will be established, by showing that the singularities of the $n$th summand of the series are not cancelled by the infinitely many remaining terms of the series.\footnote{A nice example \cite{stein} where such a cancellation does occur is
\[{z\ov1-z}={z\ov1-z^2}+{z^2\ov1-z^4}+{z^4\ov1-z^8}+\cdots+
{z^{2^n}\ov1-z^{2^{n+1}}}+\cdots.\]} We shall see that a certain derivative of the $n$th term is unbounded as $k^2$ tends to an $n$th root of unity while the same derivative of the sum of the later terms is bounded, and if it is a primitive $n$th root the same derivative of each earlier term is also bounded.

To put what we have done into some perspective, we start with a symbol
\[(1-\x)^{\a_+}\,(1-1/\x)^{\a_-}.\]
Then we introduce its $k$-deformation, times a ``nice'' function $\ps(\x)$, and consider their Toeplitz determinants as functions of the paremeter $k$ inside the unit circle.
Is it important that we begin with a symbol with only one singularity on the boundary? It is not. We may begin instead with a general Fisher-Hartwig symbol  \cite{fh}
\[\prod_{p=1}^P\,(1-u_p\x)^{\a_p^+}\ \prod_{q=1}^Q(1-v_q/\x)^{\a_q^-},\]
where $|u_p|,\,|v_q|=1$ and $P,\,Q>0$. With some conditions imposed on the 
$\a_p^+$ and the $\a_q^-$, we show that the conclusion of the theorem holds for the deformations of these symbols.

Here is an outline of the paper. In the next section we derive the expansion for $\ch(k)$ as a series of multiple integrals. In the following section the theorem is proved, and in the section after that we show how to extend the result to (almost) general Fisher-Hartwig symbols. In two appendices we give the proof of a proposition used in Section II and proved in \cite{tw2}, and discuss a minimum question that arises in Section~IV.

\bc{\bf II. Preliminaries}\ec

We invoke the formula of Geronimo-Case \cite{geronimo} and Borodin-Okounkov \cite{borodin} to write the Toeplitz determinant in terms of the Fredholm determinant of a product of Hankel operators. The Hankel operator $H_N(\ph)$ is the operator on $\l^2(\Z^+)$ with kernel $(\ph_{i+j+N+1})_{i,j\ge0}$. 

We have a factorization $\ph(\x)=\ph_{+}(\x)\,\ph_{-}(\x)$, where $\ph_+$ extends analytically inside the unit circle and $\ph_-$ outside, and $\ph_+(0)=\ph_-(\iy)=1$. More explicitly,
\[\ph_{+}(x) = (1-k\, \x)^{\a_+}\,\ps_+(\x)\>\>\>\textrm{and}\>\>\> \ph_{-}(\x) = (1 - k/\x)^{\a_-}\,\ps_-(\x).\]
If $\ps(\x)$ is analytic and nonzero for $s<|\x|<s\inv$ then 
$\ps_+(\x)$ resp. $\ps_-(\x)$ is analytic and nonzero for $|\x|<s\inv$ resp. $|\x|>s$.  

The formula of G-C/B-O is 
\[D_N(\ph)=E(\ph)\,\det\Big(I-H_N\Big({\ph_-\ov\ph_+}\Big)
\,H_N\Big({\tl\ph_+\ov\tl\ph_-}\Big)\Big),\]
where for a function $f$ we define $\tl f(\x)=f(\x\inv)$.
Thus, if we write
\[\L(\x)={\ph_-(\x)\ov\ph_+(\x)}=(1-k\, \x)^{-\a_+}\,(1-k/\x)^{\a_-}\,{\ps_-(\x)\ov\ps_+(\ps)},\]
\be K_N=H_N(\L)\,H_N(\tl\L\inv),\label{KN}\ee
then $\ch(k)$ equals $E(\ph)$ times
\[\S(k)= \sum_{N=1}^\iy\left[\det(I-K_N)-1\right]. \]

In \cite{tw2} the following was proved. We give the proof in Appendix A.
\sp

\noi{\bf Proposition}. Let $H_N(du)$ and $H_N(dv)$ be two Hankel matrices acting on $\l^2(\Z^+)$ with $i,j$ entries
\be\int x^{N+i+j}\,d u(x),\ \ \ \int y^{N+i+j}\,dv(y),\label{uv}\ee
respectively, where $u$ and $v$ are measures supported inside the unit circle. Set $K_N=H_N(du)\,H_N(dv)$. Then
\[\sum_{N=1}^\iy [\det(I-K_N)-1]\]
\[=\sum_{n=1}^\iy{(-1)^n\ov (n!)^2}
\int\cd\int{\prod_i x_iy_i\ov1-\prod_i x_iy_i}\(\det\({1\ov 1-x_iy_j}\)\)^2\,\prod_i du(x_i)\,dv(y_i),\]
where indices in the integrand run from 1 to $n$.
\sp 

We apply this to the operator $K_N=H_N(\L)\,H_N(\tl\L\inv)$ given by (\ref{KN}). The matrix for $H_N(\L)$ has $i,j$ entry
\[{1\ov2\pi i}\int\L(\x)\,\x^{-N-i-j-2}\,d\x,\]
where the integration is over the unit circle. The integration may be taken over a circle with radius in $(1,|k|\inv)$ as long as $|k|>s$. (Recall that $\ps_\pm(\x)$ are analytic and nonzero for $s<|\x|<s\inv$.) We assume this henceforth. 

Setting $\x=1/x$ we see that the entries of $H_N(\L)$ are given as in (\ref{uv}) with
\[du(x)={1\ov2\pi i}\,\L(x\inv)\,dx,\]
and integration is over a circle $\C$ with radius in $(|k|,1)$.
Similarly, $H_N(\tl\L\inv)=H_N(v)$ where in (\ref{uv})
\[dv(y)={1\ov2\pi i}\,\L(y)\inv\,dy,\]
with integration over the same circle $\C$.

Hence the Proposition gives
\be \S(k)=\sum_{n=1}^\iy \S_n(k),\label{sumS}\ee
where
\[\S_n(k)={(-1)^n\ov (n!)^2}{1\ov(2\pi i)^{2n}}
\int\cd\int{\prod_i x_iy_i\ov1-\prod_i x_iy_i}\(\det\({1\ov 1-x_iy_j}\)\)^2\,\prod_i {\L(x_i\inv)\ov \L(y_i)}\, \prod_i dx_i\,dy_i,\]
with all integrations over $\C$.

We deform each $\C$ to the circle with radius $|k|$ (after which there are integrable singularities on the contours). Then we
make the substitutions $x_i\to k x_i,\ y_i\to k y_i$, and obtain
\be\S_n(k)={(-1)^n\ov (n!)^2}\,{\k^{2n}\ov(2\pi i)^{2n}}
\,\int \cd\int {\prod_i x_iy_i\ov1-\k^n\prod_i  x_iy_i}\;\(\det\({1\ov 1-\k x_iy_j}\)\)^2\,
\prod_i {\L(k\inv x_i\inv)\ov \L(ky_i)}\,\prod_i dx_i\,dy_i,\label{Sn1}\ee
where integrations are on the unit circle. We record that
\be{\L(k\inv x\inv)\ov \L(ky)}={(1-\k x)^{\a_-}\ov (1-\k y)^{-\a_+}}\;
{(1-x\inv)^{-\a_+}\ov (1-y\inv)^{\a_-}}\;{\r(k\inv x\inv)\ov\r(ky)},\label{LaLa}\ee
where we have set
\[\k=k^2,\ \ \ \r(x)={\ps_-(x)\ov\ps_+(x)}.\] 
The complex planes are cut from $\k\inv$ to $\iy$ for the first quotient in (\ref{LaLa}) and from~0 to~1 for the second quotient.

Using the fact that the determinant in the integrand is a Cauchy determinant we obtain the alternative expression
\be \S_n(k)={(-1)^n\ov (n!)^2}\,{\k^{n(n+1)}\ov(2\pi i)^{2n}}\,\int\cd\int{\prod_i x_iy_i\ov1-\k^n \prod_i  x_iy_i}\;{\D(x)^2\,\D(y)^2\ov\prod_{i,j}(1-\k\,x_iy_j)^2}\,
\prod_i {\L(k\inv x_i\inv)\ov \L(ky_i)}\,\prod_i dx_i\,dy_i,\label{Sn2}\ee
where $\D(x)$ and $\D(y)$ are Vandermonde determinants.
\sp

For any $\d<s$ we can deform each contour of integration to one that goes back and forth along the segment $[1-\d,1]$ and then around the circle with center zero and radius $1-\d$.\footnote{To expand on this, it goes from $1-\d$ to 1 just below the interval $[1-\d,1]$, then from 1 to $1-\d$ just above the interval $[1,1-\d]$, then counterclockwise around the circle with radius $1-\d$ back to $1-\d$.} This is the contour we use from now on.

\bc{\bf III. Proof of the Theorem}\ec

There will be three lemmas. In these, $\ep\ne1$ will be an $n$th root of unity and we consider the behavior of $\S(k)$ as $\k\to\ep$ radially. Because the argument that follows involves only the local behavior of $\S(k)$, we may consider $\k$ as the underlying variable and in (\ref{LaLa}) replace $k$ by the appropriate $\sqrt{\k}$.  We define 
\[\m=\k^{-n}-1,\ \ \ \b=\a_++\a_-,\ \ \ b=\Re\b,\] 
so that $\m>0$ and $\mu\to0$ as $\k\to\ep$. 
\sp

\noi{\bf Lemma 1}. We have\footnote{We use the usual notation $[b n]$ for the greatest integer in $b n$. The symbol $\approx$ here indicates that the ratio tends to a nonzero constant as $\m\to0$.}
\[\({d\ov d\k}\)^{2n^2-[b n]}\S_n(k)\approx\m^{[b n]-\b n-1}.\]

\noi{\bf Proof.} We set
\[\l=2n^2-[b n]\]
and first consider
\be\int\cdots\int
{\prod_i x_iy_i\ov(1-\k^n \prod_i x_iy_i)^{\l+1}}\;{\D(x)^2\,\D(y)^2\ov\prod_{i,j}(1-\k\,x_iy_j)^2}\,\prod_i{\L(k\inv x_i\inv)\ov\L(ky_i)}\,\prod_i dx_i\,dy_i,\label{lint}\ee
where all indices run from 1 to $n$. This will be the main contribution to 
$d^\l \S_n(k)/d\k^\l$.
 
For the $i,j$ factor in the denominator in the second factor, if $x_i$ or $y_j$ is on the circular part of the contour then $|x_iy_j|\le 1-\d$ and the factor is bounded away from zero; otherwise $x_iy_j$ is real and positive and this factor is bounded away from zero as $\k\to\ep$ since $\ep\ne1$. So we consider the rest of the integrand.

If $\prod_i |x_iy_i|<1-\d$ then the rest of the integrand is bounded except for the last quotient, and the integral of that is $O(1)$ since $\Re\a_\pm<1$.

When $\prod_i|x_iy_i|>1-\d$ then each $|x_i|,\,|y_i|>1-\d$, so each $x_i,\,y_i$ is integrated below and above the interval $[1-\d,1]$. If all the integrals are taken over the interval itself we must multiply the result by the nonzero constant $(4\,\sin\pi\a_+\,\sin\pi\a_-)^n$. The factors $1-\k\,x_iy_j$ in the second denominator equal $1-\k(1+O(\d))=(1-\k)\,(1+O(\d))$ since $\k$ is bounded away from~1. From this we see that if we factor out $\k^{(\l+1)n}$ from the first denominator, $(1-\k)^{n^2}$ from the second denominator, and 
$(1-\k)^{\b n}(\r(k\inv)/\r(k))^n$ from the last factor (all of these having nonzero limits as $\k\to\ep$), the integrand becomes 
\[{\D(x )^2\,\D(y)^2\ov(\k^{-n}-\prod_ix_i y_i)^{\l+1}}\,
\prod_i(1-x_i)^{-\a_+}\,(1-y_i)^{-\a_-}\,(1+O(\d)).\]

We make the substitutions $x_i=1-\x_i,\,y_i=1-\e_i$ and set $r=\sum_i(\x_i+\e_i)$. Then since $\prod_i(1-\x_i)(1-\e_i)=1-r+O(r^2)$ this becomes
\[{\D(\x )^2\,\D(\e)^2\ov(\m+r+O(r^2))^{\l+1}}\,
\prod_i\x_i^{-\a_+}\,\e_i^{-\a_-}\,(1+O(\d)).\]   
The integration domain becomes $r<\d+O(\d^2)$. Consider first the integral without the $O(\d)$ term. By homogeneity of the Vandermondes and the product, the integral equals a nonzero constant\footnote{This is the integral of $\D(\x )^2\,\D(\e)^2\,\prod \x_i^{-\a_+}\,\e_i^{-\a_-}$ over $r=1$. It can be evaluated using a Selberg integral, with the result
\[{1\ov\Gamma(2n^2-\b n)}\,\prod_{j=0}^{n-1}\G(j+2)^2\,\G(j-\a_++1)\,
\G(j-\a_-+1).\]\label{Selberg}} 
times
\be\int_0^{\d+O(\d^2)}{r^{2n^2-\b n-1}\ov(\m+r+O(r^2))^{\l+1}}\,dr.\label{int1}\ee

Making the substitution $r\to\m r$ results in
\pagebreak
\be \m^{2n^2-\b n-\l-1}\int_0^{(\d+O(\d^2))/\m}\,{r^{2n^2-\b n-1}\ov(1+r+
O(\m^2 r^2))^{\l+1}}\,dr\label{int2}\ee
\[=\m^{[bn]-\b n-1}\int_0^{(\d+O(\d^2))/\m}\,{r^{2n^2-\b n-1}\ov(1+r+
O(\m^2 r^2))^{2n^2-[bn]+1}}\,dr,\]
where we have put in our value of $\l$. The integral has the $\m\to0$ limit the convergent integral
\[\int_0^\iy\({r\ov1+r}\)^{2n^2-[bn]+1}\,r^{[bn]-\b n-2}\,dr,\]
and (\ref{int1}) is asymptotically this times $\m^{[bn]-\b n-1}$. 

For the integral with the $O(\d)$ we take the absolute values inside the integrals and find that it is $O(\d)$ times what we had before, except that the $\b$ in the exponents are replaced by $b$, and in footnote~\ref{Selberg} the exponents $\a_\pm$ are replace by their real parts. Since $\d$ is arbitrarily small, it follows that the intergral of (\ref{int1}) is asymptotically a nonzero constant times $\m^{[bn]-\b n-1}$. 

To compute the derivative of order $2n^2-[bn]$ of the integral in (\ref{Sn2}) one integral we get is what we just computed. The other integrals are similar but in each the $\l$ in the first denominator is at most $2n^2-[b n]-1$, while we get extra factors obtained by differentiating the rest of the integrand for $\S_n(k)$. These factors are of the form $(1-\k x_iy_i)\inv,\ (1-\k x_i)\inv$, $(1-\k y_i)\inv$, or derivatives of $\r(k\inv x_i\inv)$ or of $\r(ky_i)\inv$. These are all bounded. Because $\l\le 2n^2-[b n]-1$ the integral (\ref{int1}) is $O(\m^{-1+\g})$ for some $\g>0$. The lemma follows.\qed
\sp

\noi{\bf Lemma 2}. If $\ep^m\ne1$ then
\[\({d\ov d\k}\)^{2n^2-[b n]}\S_m(k)=O(1).\]

\noi{\bf Proof}. If $\ep^m\ne1$ all terms, aside from those coming from the last factors, obtained by differentiating the integrand in (\ref{Sn2}) with $n$ replaced by $m$ are bounded as $\k\to\ep$. Differentiating the last factor in the integrand any number of times results in an intregrable function.\qed
\sp

\noi{\bf Lemma 3}. We have 
\[\sum_{m>n}\({d\ov d\k}\)^{2n^2-[b n]}S_m(k)=O(1).\]

\noi{\bf Proof}. We shall show that for $\k$ sufficiently close to $\ep$ all integrals we get by differentiating the integral for $S_m(k)$ are at most $A^m\,m^m$, where $A$ is some constant.\footnote{The value of $A$ will change with each of its appearances. It may depend on $n$ and $\d$, which are fixed, but not on~$m$.}  Because of the $1/(m!)^2$ appearing in front of the integrals this will show that the sum is bounded.  

As before, we first use (\ref{Sn2}) with $n$ replaced by $m$, and consider the integral we get when the first factor in the integrand is differentiated $2n^2-[b n]$ times. All indices in the integrands now run from $1$ to $m$. 

First, 
\[|1-\k^m\prod_i x_iy_i|\ge1-\prod_i |x_iy_i|.\]
Next we use that either $|x_i|=1-\d$ or $x_i\in[0,1]$, and $\k\in [0,\ep]$, 
to see that $|1-\k x_i|\ge\min(\d, d)$, where $d={\rm dist}(1,\,[0,\ep])$. We may assume $\d<d$. Then $|1-\k x_i|\ge\d$, and similiarly, $|1-\k y_i|\ge \d$. It follows that the integrand in (\ref{Sn2}) after differentiating the first factor has absolute value at most $A^m$ times
\be{1\ov(1-\prod_i |x_iy_i|)^{2n^2-[b n]+1}}\;{\D(x)^2\,\D(y)^2\ov\prod_{i,j}|1-\k x_iy_j|^2}\prod_i|1-x_i|^{-a_+}\,|1-y_i|^{-a_-},\label{integrand}\ee
where $a_\pm=\Re\a_\pm$.

If $\prod_i|x_iy_i|<1-\d$ then the first factor is at most $\d^{-2n^2+[b n]-1}$. When $\prod_i|x_iy_i|>1-\d$ we set, as before, $x_i=1-\x_i,\,y_i=1-\e_i$ with $\x_i,\,\e_i\in[0,\d]$. Since we are to integrate back and forth over these intervals we must multiply the estimate below by the irrelevant factor $2^{2m}$.  

We have 
$\prod_i (1-\x_i )(1-\e_i )\le (1-\x_i )(1-\e_i )$ for each $i$, and so averaging gives 
\[\prod_i (1-\x_i )(1-\e_i )\le{1\ov 2m}\sum_i(1-\x_i )(1-\e_i ),\]
and therefore
\[ 1-\prod_i (1-\x_i )(1-\e_i )\ge {1\ov2m}\sum_i\(1-(1-\x_i )(1-\e_i )\)\]
\be ={1\ov2m}\sum_i(\x_i+\e_i-\x_i\e_i)\ge {1\ov2m}\sum_i(\x_i+\e_i)/2\label{lowerbound}\ee
if $\d<1/2$, since each $\x_i,\e_i<\d$.
From this we see that in the region where $\sum_i(\x_i+\e_i)>\d$ the first factor in (\ref{integrand}) is at most 
$(4m/\d)^{2n^2-[bn]+1}$. 

So in either of these two regions the first factor is at most $A^m$. We then use (\ref{integrand}) with the second factor replaced by the absolute value of
\[\(\det\({1\ov 1-\k x_iy_j}\)\)^2.\]
Each denominator has absolute value at least $\d$, so by the Hadamard inequality the square of the determinant has absolute value at most $\d^{-2m}\,m^m$. Therefore the integral over this region has absolute value at most
\[A^m\,m^m\,\int\cd\int\prod_i|1-x_i|^{-a_+}\,|1-y_i|^{-a_-}\;\prod_i dx_i\,dy_i.\]
The integral here is $A^m$, and so we have shown that the integral in the described region is at most $A^m\,m^m$.

It remains to bound the integral over the region where $x_i=1-\x_i,\,y_i=1-\e_i$ with $\x_i,\e_i\in[0,\d]$, and $r=\sum_i(\x_i+\e_i)<\d$. Using (\ref{lowerbound}) again, we see that the integrand has absolute value at most $A^m$ times
\[d^{-m^2}{\D(\x)^2\,\D(\e)^2\ov (\sum_i(\x_i+\e_i))^{2n^2-[b n]+1}}\prod_i\x_i^{-a_+}\,\e_i^{-a_-}.\]
(Recall that $d={\rm dist}(1,\,[0,\ep])$, snd $\k x_iy_j\in[0,\ep]$. The factor $(4m^2)^{2n^2-[\g n]+1}$ coming from using (\ref{lowerbound}) were absorbed into $A^m$.) Integrating this with respect to $r$ over $r<\d$, using homogeneity, gives
\[\int_{r=1}\D(\x)^2\,\D(\e)^2\prod_i\x_i^{-a_+}\,\e_i^{-a_-}\,d(\x,\e)\]
(where $d(\x,\e)$ denotes the $(2n-1)$-dimensional measure on $r=1$) times
\[d^{-m^2}\int_0^\d r^{2m^2-2n^2+[b n]-b n-1}\,dr.\]
The first integral is given in footnote \ref{Selberg} with $n$ replaced by $m$ and $\a_\pm$ replaced by $a_\pm$, and is exponentially small in $m$. The last integral is $O(\d^{2m^2})$ since $m>n$ and $n$ is fixed. Since $\d^2<d$, the product is exponentially small in $m$.  

So we have obtained a bound for one term we get when we differentiate $2n^2-[\g n]$ times the integrand for $\S_m(k)$. The number of factors in the integrand involving $\k$ is $O(m^2)$ so if we differentiate $2n^2-1$ times we get a sum of $O(m^{4n^2})$ terms. In each of the other terms the denominator in the first factor has a power even less than $2n^2-[\g n]$ and at most $2n^2$ extra factors appear which are of the form $(1-\k x_iy_i)\inv,\,(1-\k x_i)\inv$, or $(1-\k y_i)\inv$. Also, $\r(k\inv x_i\inv)$ or $\r(ky_i)\inv$ may be replaced by some of its derivatives. Each has absolute value at most $\d\inv$, so their product is $O(\d^{-4n^2})$. It follows that we have the bound $A^m\,m^m$ for the sum of these integrals. Lemma~4 is established.\qed
\sp

\noi{\bf Proof of the Theorem}. Let $\ep$ be a primitive $n$th root of unity. 
Then $\ep^m\ne1$ when $m<n$ so Lemma 2 applies for these~$m$. Combining this with Lemmas~1 and 3 we obtain 
\[\({d\ov d\k}\)^{2n^2-[b n]}\S(k)\approx\m^{[b n]-\b n-1}\]
as $\k\to\ep$. This is unbounded, so $\S(k)$ cannot be analytically continued beyond any such $\ep$, and these are dense in the unit circle.

Thus the unit circle is a natural boundary for $\S(k)$, and this implies that the same is true of $\ch(k)$.\qed
\pagebreak

\bc{IV. \bf Fisher-Hartwig symbols}\ec

In this section we show how to extend the proof of the theorem to deformations of Fisher-Hartwig symbols. 

We start with a Fisher-Hartwig symbol\footnote{We could easily add a factor $\ps(\x)$ to give the general Fisher-Hartwig symbol.}
\[\prod_{p=1}^P\,(1-u_p\x)^{\a_p^+}\,\prod_{q=1}^Q(1-v_q/\x)^{\a_q^-},\]
where $|u_p|,\,|v_q|=1$ and $P,\,Q>0$, and then its $k$-deformation\[\ph(\x)=\prod_{p=1}^P\,(1-ku_p\x)^{\a_p^+}\,\prod_{q=1}^Q(1-kv_q/\x)^{\a_q^-}.\] 
We assume that $\Re\a_p^+,\ \Re\a_q^-<1$ and $\a_p^+,\,\a_q^-\not\in\Z$. (Plus a simplifying assumption that comes later.)

The singularities of $D_N(\ph)$ on the unit circle are at the $(u_p v_q)^{-1/2}$, and
\[E(\ph)=\prod_{p,q}(1-k^2 u_p v_q)^{-\a_p^+\a_q^-}.\]

We have now
\[\L(\x)=\prod_{p,q}(1-ku_p\x)^{-\a_p^+}\,(1-kv_q/\x)^{\a_q^-},\]
\[{\L(k\inv x\inv)\ov \L(ky)}=\prod_{p,q}{(1-u_p/x)^{-\a_p^+}\,
(1-\k v_q x)^{\a_q^-}\ov(1-\k u_p y)^{-\a_p^+}\,(1-v_q/y)^{\a_q^-}}.\]

Again we begin by considering the integral
\be\int\cdots\int
{\prod_i x_iy_i\ov(1-\k^n \prod_i x_iy_i)^{\l+1}}\;{\D(x)^2\,\D(y)^2\ov\prod_{i,j}(1-\k\,x_iy_j)^2}\,\prod_i{\L(k\inv x_i\inv)\ov\L(ky_i)}\,\prod_i dx_i\,dy_i.\label{int3}\ee

Our integrations are for the $x_i$ around the cuts $[1-\d,1]\,u_p$ and for the $y_i$ around the cuts $[1-\d,1]\,v_q$ and then both around the circle with radius $1-\d$. (In case we do want to generalize with a factor $\ps(\x)$ as before.) If we replace integrals around the cuts by integrals on the cuts, then for a cut $[1-\d,1]\,u_p$ we must multiply by $2\,\sin\pi\a_p^+$ and for a cut $[1-\d,1]\,v_q$ we multiply by $2\,\sin\pi\a_q^-$. (These are both nonzero.) We assume that this has been done.

We now let $\k\to\ep$ radially, where $\ep$ is an $n$th root of $\prod(u_{p_i}\,v_{q_i})\inv$, but not equal to any $(u_p v_q)\inv$.  We also choose it so that it is not an $m$th root of any product of the form $\prod (u_{p_i}\,v_{q_i})\inv$ with $m<n$ . These $\ep$ become dense on the unit circle as $n\to\iy$. The last condition assures that the integrals with $m<n$ are bounded, which will give the analogue of Lemma 2. We now consider the analogue of Lemma 1.

The integral over $\prod|x_iy_i|<1-\d$ is bounded, as before. In the region where $\prod|x_iy_i|>1-\d$ each $x_i$ and $y_i$ is integrated on the union of its associated cuts. This is the sum of integrals in each of which each $x_i$ is integrated over one of the cuts and each $y_i$ is integrated over of the cuts. Suppose that $x_i$ is integrated over $[1-\d,1]\,u_{p_i}$ and $y_i$ is integrated over $[1-\d,1]\,v_{q_i}$. (We consider this one possibility at first. Then we will have to sum over all possibilities.) 

If we factor out $\prod u_{p_i}v_{q_i}$ from the first numerator, $\prod(1-\k u_{p_i}v_{q_i})^2$ from the second denominator, and 
$\prod(1-\k\,u_{p_i}v_{q_i})^{\a_{p_i}+\a_{p_i}}$ from the last product the integrand becomes\linebreak $1+O(\d)$ times
\be{\D(x )^2\,\D(y)^2\ov(1-\k^{n}\prod_ix_i y_i)^{\l+1}}\,
\prod_i(1-u_{p_i}/x_i)^{-\a_{p_i}^+}\,(1-v_{q_i}/y_i)^{-\a_{q_i}^-}.\label{D0}\ee

We make the substitutions $x_i=(1-\x_i)\,u_{p_i},\ y_i=(1-\e_i)\,v_{q_i}$, and define 
\[I_p=\{i:p_i=p\},\ \ I_q=\{i:q_i=q\}.\]
Then 
\[\prod_i(1-u_{p_i}/x_i)^{-\a_{p_i}^+}\,(1-v_{q_i}/y_i)^{-\a_{q_i}^-}=\prod_{p,\,i\in I_p}\x_i^{-\a_p^+}\,\cdot\,\prod_{q,\,i\in I_q}\e_i^{-\a_q^-}\times (1+O(\d)).\]

As for the Vandermondes, we have 
\be\D(x)=\pm\prod_p\,\D(x_i:i\in I_p)\,\cdot\,\prod_{p\ne p'}\ \prod_{j\in I_p,\,j'\in I_{p'},\,j<j'}(x_j-x_{j'}),\label{D1}\ee
and similarly for $\D(y)$. If we define $n_p=|I_p|,\ n_q=|I_q|$, then the last double product is to within a factor $1+O(\d)$ equal to
\[\pm\,\prod_{p<p'}(u_p-u_{p'})^{n_p n_{p'}},\]
while the first product is to within a factor $1+O(\d)$ equal to
\[\prod_p u_p^{n_p(n_p+1)/2}\,\prod_p\,\D(\x_i:i\in I_p).\]

Thus, if we factor out $(\k^{n}\prod u_{p_i}\,v_{q_i})^{\l+1}$ from the denominator in (\ref{D0}), and set $\m=\k^{-n}\prod (u_{p_i}\,v_{q_i})\inv-1$, then (\ref{D0}) may get replaced by a constant times $1+O(\d)$ times
\be{1\ov(\m+\sum_i(\x_i+\e_i))^{\l+1}}\prod_{p,\,i\in I_p}\D(\x_i:i\in I_p)^2\,\x_i^{-\a_p^+}\,\cdot\,\prod_{q,\,i\in I_q}\D(\e_i:i\in I_q)^2\,
\e_i^{-\a_q^-}.\label{D4}\ee

If we use homogeneity the integral of (\ref{D4})  becomes a nonzero constant\footnote{Also computable using a Selberg integral, it is 
\[{1\ov\Gamma(\sum_p n_p(n_p-\a_p^+)+\sum_q n_q(n_q-\a_p^-))}\,\prod_p\prod_{j=0}^{n_p-1}\G(j+2)\,\G(j-\a_p^++1)\,\cdot\,\prod_q\prod_{j=0}^{n_q-1}\G(j+2)\,\G(j-\a_q^-+1).\]\label{selberg}} 
times
\be\int_0^{\d}{1\ov(\m+r)^{\l+1}}\,r^{-1+\sum_p n_p(n_p-\a_p^+)+\sum_q n_q(n_q-\a_q^-)}\,dr.\label{int4}\ee
This is largest when the power of $r$ is smallest. So we minimize 
\[\sum_p n_p(n_p-a_p^+),\ \ \ (a_p^+=\Re\a_p^+)\]
over all $\{n_p\}$ with $n_p\ge0,\ \sum_p n_p=n$. The solution are not necessarily unique. But in any case
\be M_n^+:=\min\sum_p n_p(n_p-a_p^+)={n^2\ov P}+O(n),\label{Mnp}\ee
and $M_{n+1}^+>M_{n}^+$ for large enough $n$.\footnote{See Appdendix B.} 

Similarly, with $a_q^-=\Re\a_q^-$ and
\be M_n^-:=\min\sum_q n_q(n_q-a_q^-)={n^2\ov Q}+O(n).\label{Mnm}\ee
 
Then we choose 
\be\l=\sum_p n_p^2+\sum_q n_q^2-\Big[\sum_p n_p\,a_p^++\sum_q n_q\,\ a_q^-\Big]\label{ell}\ee
with the minimal $n_p$ and $n_q$.
The integral (\ref{int4}) is equal to
\[\m^{-1+[\sum_p n_p\,a_p^++\sum_q n_q\,\ a_q^-]-(\sum_p n_p\,\a_p^++\sum_q n_q\,\a_q^-)}\]
times
\[\int_0^{\d/\m}{1\ov(1+r)^{\l+1}}\,r^{-1+\sum_p n_p(n_p-\a_p^+)+\sum_q n_q(n_q-\a_q^-)}\,dr.\]
The exponent of $\m$ has real part in $(-2,-1]$ and the integral has a nonzero limit (a Beta function) as $\m\to0$. 

Once we take care of the integrals with the $O(\d)$ as in the proof of Lemma 1 we deduce that this is the asymptotic result for the integral when we choose this set of cuts. 

{\it We now assume the minimal solutions are unique.}\footnote{We shall see in Appendix B that for large $n$ uniqueness is a condition on the $a_p^+$ and $a_q^-$ that depends only on the residue classes of $n$ modulo $P$ and $Q$. It suffices for our purposes that we have uniqueness for some sequence $n\to\iy$.} 

Then for the other choices of cuts the integral (\ref{int4}) is $O(\m^{-1+\gamma})$ for some $\gamma>0$, and so the integral over the chosen set of cuts dominates. We still have to allocate the $x_i$ and $y_i$ to the various cuts, once the numbers of each have been chosen. The number of ways of doing this is $n!/\prod n_p!$ for the~$x_i$ and $n!/\prod n_q!$ for the $y_i$. (The total number of ways is at most $P^n\,Q^n$.) 

This takes care of the integral (\ref{int3}), the main contributions to 
$(d/d\k)^\l S_n(\k)$. We complete the proof of the analogue of Lemma 1 as we did at the end of the proof of that lemma.
Thus, with $\l$ given by (\ref{ell}),
\[\({d\ov d\k}\)^\l\,\S_n(k)\approx\m^{-1+[\sum_p n_p\,a_p^++\sum_q n_q\,\ a_q^-]-(\sum_p n_p\,\a_p^++\sum_q n_q\,\a_q^-)}.\]

For the analogue of Lemma 3 we first consider the integral (\ref{int3}) with $n$ replaced by $m>n$, and $\l$ given by (\ref{ell}). As before it remains to bound the integrals over the regions where each $x_i=(1-\x_i)u_{p_i}$ and each $y_i=(1-\e_i)v_{q_i}$, with $\x_i,\e_i\in[0,\d]$, and $r=\sum_i(\x_i+\e_i)<~\d$. 

Replacing the first denominator in (\ref{int3}) by $(\sum(\x_i+\e_i))^{\l+1}$ introduces a factor $(4m^2/\d)^{\l+1}$ as before, a factor that can be ignored. The reciprocal of the second denominator is at most $d^{-m^2}$ where $d=\min_{p,q}{\rm dist}([0,\ep],(u_pv_q)\inv)$. The product of the terms involving $\k$ in the last product is $d^{-O(m)}$, and so may also be ignored. The square of the product over $p<p'$ in (\ref{D1}), times the square of the analogous product over $q<q'$, is at most $2^{(P^2+Q^2)m^2}$.
There remains an integrand whose absolute value is bounded by
\[{1\ov(\sum_i(\x_i+\e_i))^{\l+1}}\prod_{p,\,i\in I_p}\D(\x_i:i\in I_p)^2\,\x_i^{-a_p^+}\,\cdot\,\prod_{q,\,i\in I_q}\D(\e_i:i\in I_q)^2\,
\e_i^{-a_q^-}.\]

The integral of the products over $r=1$ (given exactly in footnote \ref{selberg}) is trivially at most its maximum (at most $A^m\,4^{m^2}$) times the $(2m-1)$-dimensional measure of $r=1$, which is $1/\G(2m)$. We use the crude bound $4^{m^2}$. This is to multiply
\[\int_0^{\d}r^{-\l-2+\sum_p m_p(m_p-a_p^+)+\sum_q m_q(m_q-a_q^-)}\,dr.\]
Now 
\[\sum_p m_p(m_p-a_p^+)+\sum_q m_q(m_q-a_q^-)\]
is at least $M_m^++M_m^-$, and it follows from (\ref{Mnp}) and (\ref{Mnm}), and the strict monotonicity of the sequences $\{M_n^\pm\}$, that for large enough $n$ and some $R$ this greater than $\l+1+m^2/R$ for all $m>n$. Then the integral is at most $\d^{m^2/R}$. 

This integral is one of at most $P^mQ^m$ integrals, and this factor also can be ignored. The factors we had before that could not be ignored combine to $(4\,2^{P^2+Q^2}/d)^{m^2}$.  It follows that if we choose $\d<(4\,2^{P^2+Q^2}/d)^{-R}$ the integral over $r<\d$ of (\ref{int3}) with $m$ replacing $n$ is exponentially small. 

This takes care of the integral (\ref{int3}) with $m$ replacing $n$, the main contributions to 
$(d/d\k)^\l S_m(\k)$. We complete the proof of the analogue of Lemma 3 as we did at the end of the proof of that lemma. \qed

\bc{\bf Appendix A. Proof of the Proposition}\ec

The Fredholm expansion is
\[\det(I-K_N)=1+\sum_{n=1}^\iy{(-1)^n\ov n!}\sum_{p_1,\ld,p_n\ge0}\det(K_N(p_i,p_j)).\]
Therefore its suffices to show that
\[\sum_{N=1}^\iy\,\sum_{p_1,\ld,p_n\ge0}\det(K_N(p_i,p_j))\]
\[={1\ov n!}\int\cd\int{\prod_i x_iy_i\ov1-\prod_i x_iy_i}\(\det\({1\ov 1-x_iy_j}\)\)^2\,du(x_1)\cd du(x_n)\,dv(y_1)\cd dv(y_n).\]
We have
\[K_N(p_i,p_j)=\int\int{x^{N+p_i}\,y^{N+p_j}\ov1-xy}\,du(x)\,dv(y).\]
It follows by a general identity \cite{andreief} (eqn.~(1.3) in \cite{tw1}) that
\[\det(K_N(p_i,p_j))={1\ov n!}\int\cd\int \det(x_i^{N+p_j})\,
\det(y_i^{N+p_j})\,\prod_i{1\ov 1-x_iy_i}\,\prod_i du(x_i)\,dv(y_i)\]
\[={1\ov n!}\int\cd\int \Big(\prod_i x_iy_i\Big)^N\,\det(x_i^{p_j})\,\det(y_i^{p_j})\,\prod_i{1\ov 1-x_iy_i}\,
\prod_i du(x_i)\,dv(y_i).\]
Summing over $N$ gives
\[\sum_{N=1}^\iy\,\det(K_N(p_i,p_j))=\]
\[{1\ov n!}\int\cd\int {\prod_i x_iy_i\ov 1-\prod_i x_iy_i}\,\det(x_i^{p_j})\,\det(y_i^{p_j})\,\prod_i{1\ov 1-x_iy_i}\,\prod_i du(x_i)\,dv(y_i).\]
(Interchanging the sum with the integral is justified since the supports of $u$ and $v$ are inside the unit circle.)

Now we sum over $p_1,\ld,p_n\ge0$. Using the general identity again (but in the other direction) gives
\[\sum_{p_1,\ld,p_n\ge0}\det(x_i^{p_j})\,\det(y_i^{p_j})=
n!\,\det\(\sum_{p\ge0}x_i^p\,y_j^p\)=n!\,\det\({1\ov 1-x_iy_j}\).\]

We almost obtained the desired result. It remain to show that
\be\det\({1\ov 1-x_iy_j}\)\,\prod_i{1\ov 1-x_iy_i},\label{notquite}\ee
which we obtain in the integrand, may be replaced by
\be{1\ov n!}\,\(\det\({1\ov 1-x_iy_j}\)\)^2.\label{replace}\ee
This follows by symmetrization over the $x_i$. (The rest of the integrand is symmetric.) For a permutation $\pi$, replacing the $x_i$ by $x_{\pi(i)}$ multiplies the determinant in (\ref{notquite}) by sgn\,$\pi$, so to symmetrize we replace the other factor by
\[{1\ov n!}\sum_\pi {\rm sgn}\,\pi\,{1\ov 1-x_{\pi(i)}y_i}={1\ov n!}\,\det\({1\ov 1-x_iy_j}\).\]
Thus, symmetrizing (\ref{notquite}) gives (\ref{replace}).\qed 

\bc{\bf Appendix B. The minimum question}\ec

Changing notation, we consider 
\[M_n=\min\Big\{\sum_{i=1}^k n_i\,(n_i-a_i):n_i\in\Z^+,\; \sum_{i=1}^k n_i=n\Big\},\]
and ask when this is uniquely attained. Set 
\[s=k\inv\sum_{i=1}^k a_i,\ \ \ \ab_i=(a_i-s)/2,\ \ \ \nb_i=n_i-n/k,\] 
and define
\[\N^k=\Big\{(x_i)\in\R^k:\sum_{i=1}^k x_i=0\Big\}.\]
Then $\ab=(\ab_i)\in\N^k$ and $\nb=(\nb_i)\in\N^k$. If $n\equiv\n\ ({\rm mod}\,k)$ the other conditions on the $\nb_i$ become
\[\nb_i\ge -n/k,\ \ \ \nb_i\in\Z-\n/k.\]
(Think of $\n$ as fixed and $n$ as variable.) A little algebra gives
\[\sum_{i=1}^k n_i\,(n_i-a_i)=\sum_{i=1}^k (\nb_i-\ab_i)^2+k\,(n/k-s/2)^2-\sum_{i=1}^k a_i^2/4.\]

Minimizing the sum on the left is the same as minimizing the first sum on the right, with the stated conditions on the $\nb_i$. Several things follow from this. First, since the minimum of the first sum on the right is clearly $O(1)$, the condition $\nb_i\ge -n/k$ may be dropped when $n$ is sufficiently large; second,
$M_n=n^2/k-sn+O(1)$; third (from this), $M_{n+1}-M_n=2n/k+O(1)>0$ for sufficiently large $n$; and fourth, for uniqueness we may replace our minimim problem by
\[\min\Big\{\sum_{i=1}^k (\nb_i-\ab_i)^2:\nb\in\N^k,\ \nb_i\in \Z-\n/k\Big\}.\]
This minimum is uniquely attained if and only if there is a unique point closest to $\ab$ in the set of lattice points $(\Z-\n/k)^k$ in $\N^k$. This condition depends only on the residue class of $n$ modulo $k$. 

When $k=2$ the subspace $\N^2$ is the line $x_1+x_2=0$ in $\mathbb R^2$. 
When $n$ is even the lattice consists of the points on the line with coordinates in $\Z$ and $\ab$ is equidistant from two adjacent ones when 
$a_1-a_2\in 4\Z+2$; when $n$ is odd the lattice consists of the points of the line with coordinates in $\Z+1/2$ and $\ab$ is equidistant from two adjacent ones when $a_1-a_2\in 4\Z$. Non-uniqueness occurs in these cases.
\sp

\begin{center}{\bf Acknowledgments}\end{center} 

This work was supported by the National Science Foundation through grants DMS--1207995 (first author) and DMS--1400248 (second author).

\end{document}